\documentclass{PoS}

\usepackage{amssymb, amsmath}  % math
\pdfoutput=1

\newcommand{\bi}{\begin{itemize}}
\newcommand{\ei}{\end{itemize}}

\newcommand{\be}{\begin{equation}}
\newcommand{\ee}{\end{equation}}

\newcommand{\bea}{\begin{eqnarray}}
\newcommand{\eea}{\end{eqnarray}}

\newcommand{\bt}[1]{\begin{table}[!t] \begin{center}\begin{tabular}{#1} \hline\hline  \\[-0.5em]}
\newcommand{\et}[2]{\hline\hline \end{tabular} \end{center} \caption{#1} \label{#2} \end{table}}

\def\cpt{\raise0.4ex\hbox{$\chi$}PT}
\def\scpt{S\raise0.4ex\hbox{$\chi$}PT}
\def\rscpt{rS\raise0.4ex\hbox{$\chi$}PT}
\newcommand{\EM}{electromagnetic}

\newcommand{\deltae}{\ensuremath{\Delta_E}}

\newcommand{\dem}{ \ensuremath{ \tilde{\Delta}_{\rm EM}} }

\title{Electromagnetic Corrections in Staggered Chiral Perturbation Theory}

\ShortTitle{Electromagnetic Corrections in Staggered Chiral Perturbation Theory}

\author{C.~Bernard and \speaker{E.D.~Freeland}  \\
	Department of Physics, Washington University, St.~Louis, Missouri 63130, USA\\
	E-mail: \email{edf@illinois.edu}}

\abstract{ To reduce errors in light-quark mass determinations, it is now necessary to consider electromagnetic contributions to light-meson masses. 
Calculations using staggered quarks and quenched photons are currently underway.
Suitably-extended chiral perturbation theory is necessary to extrapolate the lattice data to the physical limit.  
Here we give (preliminary) results for light-meson masses using staggered chiral perturbation theory including electromagnetism, and discuss the extent to which quenched-photon simulations 
can improve quark-mass calculations.}

\FullConference{The XXVIII International Symposium on Lattice Field Theory\\
		 June 14-19, 2010\\
		 Villasimius, Sardinia Italy}

\begin{document}

% INTRODUCTION
\section{Introduction}		\label{sec:intro}

As fundamental parameters of the Standard Model, quark masses are important to high-energy phenomenology.
Lattice gauge theory can be used for precision, non-perturbative calculations of these masses,
with state-of-the-art calculations using a vacuum including three or even four light, dynamical quarks: $u, d, s,$ and $c$.
In addition, some recent calculations have included electromagnetic effects via a quenched photon field.
This has been done, for example, by the RBC Collaboration~\cite{Blum:2010ym}.   
In calculations of quark masses by the MILC Collaboration \cite{MILC_quarkmass}, however, 
\EM\ effects have in general been
estimated based on continuum calculations.
For the calculation in Ref.~\cite{MILC_quarkmass}, the uncertainty due to electromagnetic effects dominates $m_u/m_d$ and is significant for $m_u$ and $m_d$, as shown in Table~\ref{tbl:quark_masses}. 
To put this error on better footing and to gain the higher precision desired by phenomenologists, the MILC collaboration is working to reduce the electromagnetic uncertainty in $m_u/m_d$~\cite{Torok,Basak:2008na}.
The ``rooted staggered chiral perturbation theory'' (\rscpt) expressions reported in this work are a necessary step towards that goal.  Our calculations do not assume quenched photons but
allow the sea quark charges to take nonvanishing values.  For comparison with simulations with
quenched electromagnetism one can simply set the sea quark charges to zero.

% ELECTROMAGNETIC EFFECS AND QUARK MASSES
\section{Electromagnetism and Quark Masses}

Quark masses are obtained by tuning bare-quark masses such that lattice-calculated meson masses match experiment.
Taking the $K^+$ mass as an example, one tunes the QCD-only lattice result $(M_{K^+}^2)_{\rm QCD}$ to match experiment with electromagnetic effects subtracted off~\cite{MILC_quarkmass};
\be
	(M_{K^+}^2)_{\rm QCD} \equiv M_{K^+}^2 - (M_{K^\pm}^2 - M_{K^0}^2)_{\rm EM}.  \label{eq:kaon_mass}
\ee
where $M_{K^+}^2$ is the experimental value, and the subscript ``EM'' denotes the purely \EM\ contribution\footnote{We assume, for now, that the \EM\ effects are small for $M_{K^0}^2$.}.

To obtain a value for $(M_{K^\pm}^2 - M_{K^0}^2)_{\rm EM}$, one uses an observation made by Dashen in the late 1960's
that the leading order \EM\ effects are identical in the kaon and pion systems~\cite{Dashen:1969eg}.  Specifically,
\be	
	(M_{K^\pm}^2 - M_{K^0}^2)_{\rm EM}  =  (M_{\pi^\pm}^2 - M_{\pi^0}^2)_{\rm EM}  .   \label{eq:Dashen}
\ee
Corrections to Eq.~(\ref{eq:Dashen}) are referred to as the ``violation of Dashen's theorem'' and are often parametrized as
\be
	(M_{K^\pm}^2 - M_{K^0}^2)_{\rm EM} = (1 + \Delta_E)(M_{\pi^\pm}^2 - M_{\pi^0}^2)_{\rm EM} .   \label{eq:delE}  
\ee
A lattice calculation of the ratio $(M_{K^\pm}^2 - M_{K^0}^2)_{\rm EM} / (M_{\pi^\pm}^2 - M_{\pi^0}^2)_{\rm EM}$ 
yields a value of $1+\deltae$ in which common systematics cancel. 
This can be combined with experimental measurements of $(M_{\pi^+}^2 - M_{\pi^0}^2)_{\rm EM}$ to obtain the kaon splitting needed for Eq.(~\ref{eq:kaon_mass}).

In Ref.~\cite{MILC_quarkmass}, \deltae\ was based on estimates from continuum calculations.  A direct lattice calculation of \deltae\ will allow for improved accuracy and precision.
Simulations by the MILC collaboration using ``asqtad'' staggered quarks~\cite{stag_fermion} are reported in these proceedings~\cite{Torok}.  To complete the calculation, the \rscpt\ calculation for the meson mass must also be done.  We report our preliminary results for that calculation \cite{CB_EDF} here.

% Quark mass table
\bt{l     l  l  l}
					&  $m_u$		& $m_d$		& $m_u/m_d$	\\
	central value		&  1.9		&  4.6		&  0.42		\\
	statistics			&  0.0		&  0.0		&  0.00		\\
	lattice-systematic	&  0.1		&  0.2		&  0.01		\\
	perturbative		&  0.1		&  0.2		&  --			\\
	{\bf electromagnetic} &  {\bf 0.1}	&  {\bf 0.1}	&  {\bf 0.04}		\\
\et{Quark masses as given in Ref.~\cite{MILC_quarkmass}.
Electromagnetic effects are the leading error in the ratio $m_u/m_d$ and significant in $m_u$ and $m_d$.
Electromagnetic errors are estimated from continuum calculations.  
Details about this and the determination of other errors can be found in Ref.~\protect\cite{MILC_quarkmass} 
}{tbl:quark_masses}

% THE NEUTRAL PION: PI vs PI-PRIME
\subsection{The Neutral Pion}   \label{sec:pi_piprime}

The true $\pi^0$ is a $\bar{u}u - \bar{d}d$ state.
Because of electromagnetic disconnected-diagram contributions to its mass, even for $m_u = m_d$, it is difficult to simulate.
Instead, one simulates a meson created from a light quark and a light anti-quark that are distinct flavors and have opposite electric charges~\cite{Blum:2010ym,Torok}.
We call this neutral meson the $\pi'$, and make the replacement
\be
	M_{\pi^\pm}^2 - M_{\pi^0}^2 \to M_{\pi^\pm}^2 - M_{\pi'}^2 
\ee
for the mass splitting.  
In the continuum, this approximation neglects some electromagnetic contributions to the physical $\pi^0$ 
\be
	\left(M_{\pi^0}^2 - M_{\pi'}^2 \right)_{\rm EM} 
			=  \frac{2 \dem}{16 \pi^2 f^2} M_\pi^2 \left(\ln M_\pi^2 +1 \right) + e^2 M_{\pi}^2 \: {\rm (LECs)} \ ,
\label{eq:pip}
\ee
where ``LECs'' stands for a linear combination of the low energy constants.
At the physical point, the difference given in Eq.~(\ref{eq:pip}) is small compared to the electromagnetic correction to the charged pion,
$\delta^{\rm EM} M_{\pi^\pm}^2$, which is dominated by the kaon mass:
 \be
 	\delta^{\rm EM} M_{\pi^\pm}^2 \approx  \frac{2 \dem}{16 \pi^2 f^2}  M_K^2 \ln M_K^2 + e^2 M_{K}^2 \: {\rm (LECs)} \ ,
 \ee
with
\be
	\dem \equiv \frac{4e^2C}{f^2} ,
	\label{eq:dem}
\ee
where $e$ is the fundamental electric charge, and $C$ is a leading order (LO)  LEC.

% OVERVIEW OF THE CALCULATION
\section{Overview of the Calculation}
We split the overview into two parts.  Section~\ref{sec:support} reviews the needed results from previous calculations, while Sec.~\ref{sec:schiptem} describes new work.

% SUPPORTING CALCULATIONS
\subsection{Previous Calculations}	\label{sec:support}

% Continuum		Continuum		Continuum		Continuum		Continuum
The leading-order, continuum, chiral Lagrangian including electromagnetic effects, but dropping the pure gauge-field terms, is~\cite{Urech:1994hd}
\bea
\mathcal{L}^{(2)} =	 \frac{1}{8} f^2 \langle d^\mu \Sigma^\dag d_\mu \Sigma \rangle
				+ \frac{1}{8} f^2 \langle \chi^\dag \Sigma + \chi \Sigma^\dag \rangle
				-  \frac{1}{24} m_0^2 \langle \Phi \rangle^2 
				+ e^2C \langle Q \Sigma Q \Sigma^\dag \rangle ,
				\label{eq:Lagrangian_cont}
\eea
where we work in Minkowski space,
for ease of comparison with Ref.~\cite{Urech:1994hd}.
Here, $f \approx 130$~MeV is the leading-order pion decay constant in the chiral limit,
and $\chi = 2 B_0 \mathcal{M}$, where $\mathcal{M}$ is the usual quark-mass spurion.

The first two terms in Eq.~(\ref{eq:Lagrangian_cont}) are the standard kinetic and mass terms.
The meson fields are contained in 
\be	
	\Sigma = \exp (i 2 \phi / f),
	\qquad
	\phi = 
		\begin{pmatrix}
			U	& \pi^+	& K^+ \\
			\pi^-	& D		& K^0 \\
			K^-	& \bar{K}^0	& S
		\end{pmatrix},
		\label{eq:sigma}
\ee
where diagonal entries of $\phi$ are comprised of $u, d, s$ quark--anti-quark pairs; e.g. $U = \bar{u} u$.

% eta prime and the anomaly
The anomaly term $ \frac{1}{24} m_0^2 \langle \Phi \rangle^2 $ gives mass to the $\eta'$.
It also causes the ``mesons'' on the diagonal of $\Phi$ to have disconnected propagators~\cite{Bernard:1993sv,Sharpe:2000bc}.
Usually, one takes $m_0 \to \infty$ at the start, decoupling the $\eta'$.
For a partially quenched and/or staggered calculation, though, it is more convenient to
keep the anomaly term and use the simple $U$, $D$, $S$ basis along the diagonal of $\Phi$.
In the end, we do take $m_0 \to \infty$ \cite{Sharpe:2001fh}.

% electromagnetic effects
Electromagnetic effects enter via the  the photon field $A_\mu$ in the covariant derivative \footnote{Vector and axial vector current sources have been set to zero, since they are not needed here.}
\be
	d_\mu \Sigma = \partial_\mu \Sigma - i eQ A_\mu \Sigma + i \Sigma eQ A_\mu.
	\label{eq:covderiv}
\ee
and the term $e^2C \langle Q \Sigma Q \Sigma ^\dag \rangle$, where $C$ is determined in terms of the pion splittings
via Eq.~(\ref{eq:dem}).
$Q$  is the quark (electric) charge matrix
\be
Q = \rm{diag}(q_u, q_d, q_s)
\ee
and has the property $\langle Q \rangle = 0$.
With ${\rm dim}(eQ) = {\rm dim}(p)$, $\rm{dim}(A_\mu) = 1$, and
$e \sim p$, the overall power-counting scheme is $p^2 \sim M^2 \sim m \sim e^2$.

% partially-quenched		partially-quenched		partially-quenched		partially-quenched
For lattice calculations it is useful to consider the partially quenched case where
valence and sea quarks are distinct, and may have different mass.
Taking the limit where valence and sea quark masses are equal, known as full QCD, recovers
usual chiral perturbation theory (\cpt) results.
We refer the reader to the literature~\cite{Bernard:1993sv,Sharpe:2000bc,Damgaard:2000gh} for
details.
For studying electromagnetic effects, 
the partially-quenched structure allows one to separate valence and sea quark charge contributions in an advantageous way.
The partially-quenched \cpt\ calculation including electromagnetism has been done by Bijnens and Danielsson~\cite{Bijnens:2006mk}.
This is extremely useful and provides us with cross-checks for the staggered calculation.

%  staggered no EM		staggered no EM		staggered no EM		staggered no EM		staggered no EM
The \rscpt\ calculation for meson masses done by Aubin and Bernard~\cite{Aubin:2003mg}, without electromagnetism, addresses the complexities confronted when dealing with staggered quarks.
For each continuum quark simulated, there are four staggered quarks.
These are distinguished by a quantum number referred to as ``taste".
When combined to form mesons, the result is a set of sixteen mesons of different tastes, which, at $O(a^2)$ in the lattice spacing $a$, can be grouped into five irreducible representations labeled: pseudoscalar, axial-vector, tensor, vector, singlet. 
The pseudoscalar-taste meson is a true Goldstone boson, and for that reason is
the valence meson that is usually simulated.  Here, we
calculate the mass of the Goldstone meson in \rscpt.
The use of staggered quarks brings a new, taste-violating term into the chiral Lagrangian~\cite{Aubin:2003mg}.
The two main effects of this potential at the one-loop order are the addition of taste breaking corrections to the (staggered) meson masses, and the appearance of both a taste-dependent, disconnected vertex and a taste-dependent disconnected propagator~\cite{Aubin:2003mg}.

% THE STAGGERED CHIPT WITH EM CALCULATION
\subsection{Staggered Chiral Perturbation Theory with Electromagnetic Contributions}   \label{sec:schiptem}

\begin{figure}[t!]
	\begin{center}
	\begin{tabular}{c c c c }
		\includegraphics[scale=0.1]{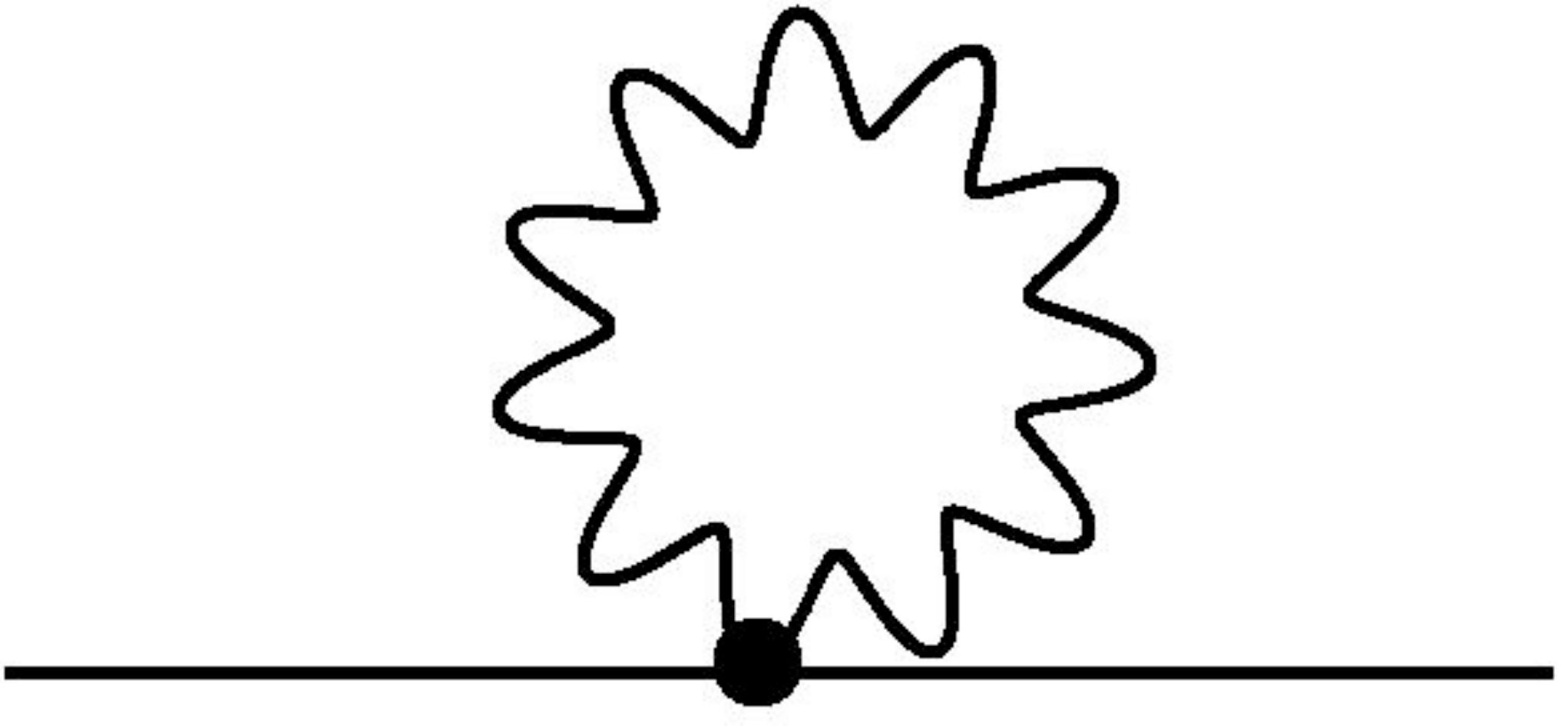}   % Photon Tadpole
		&
		\includegraphics[scale=0.1]{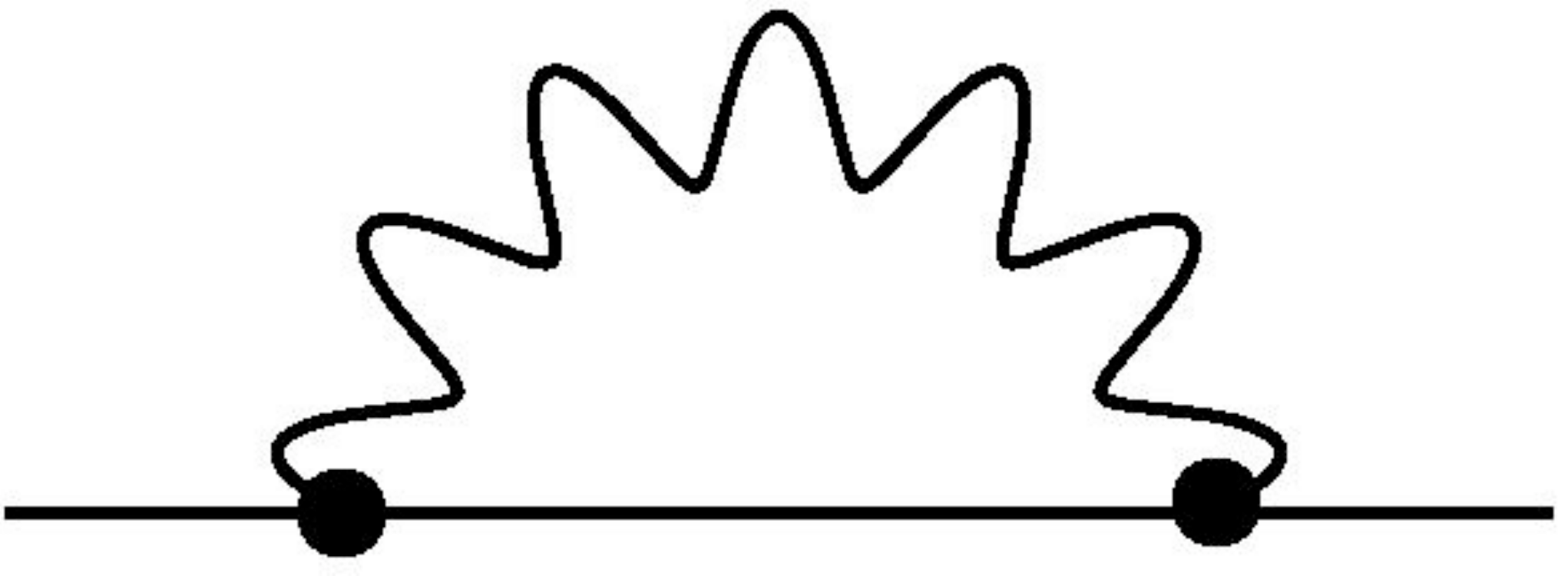}   % Photon Sunset
		&
		\includegraphics[scale=0.1]{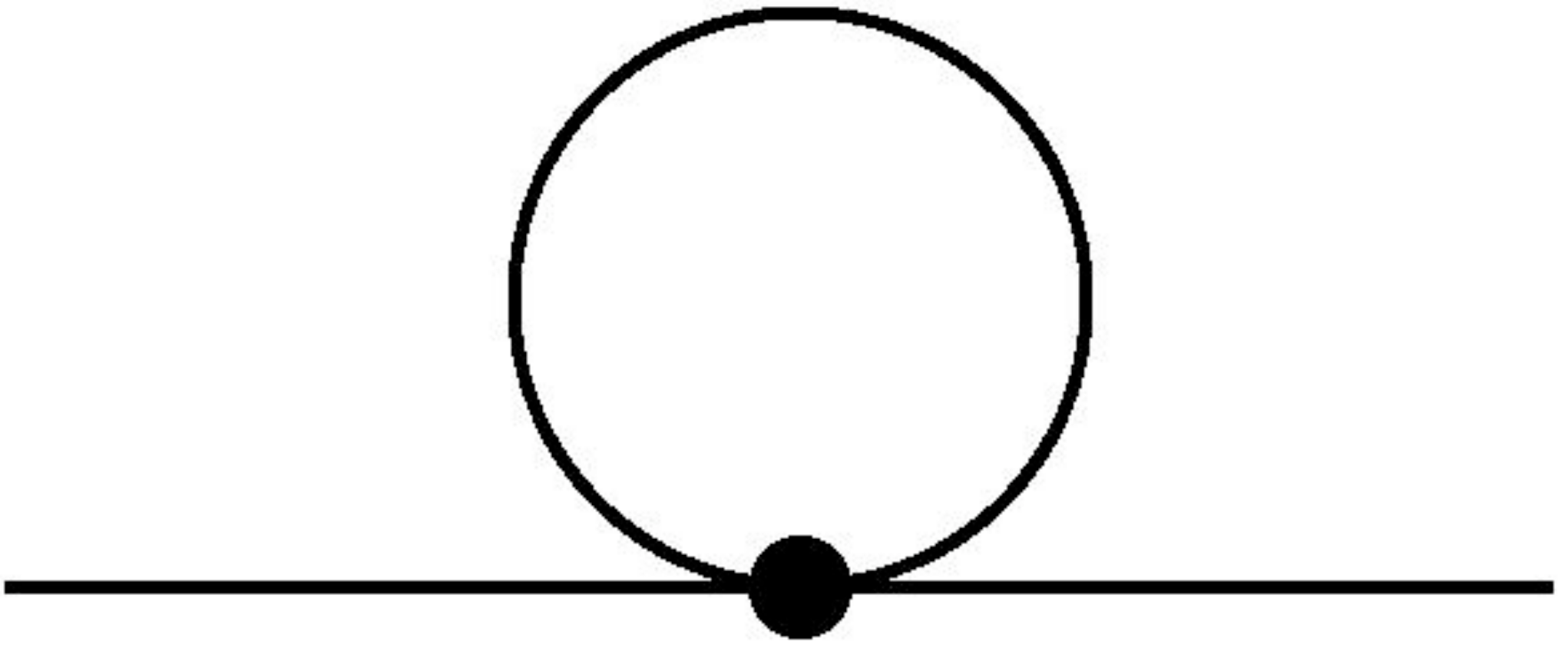}   % Meson Tadpole
		&
		\includegraphics[scale=0.1]{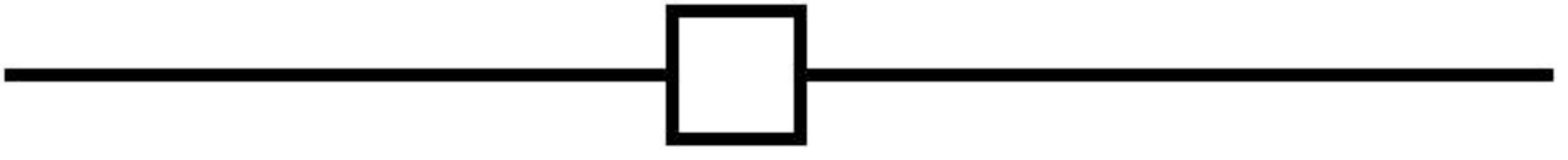}   % Tree-level Insertion
		 \\
		 (a)  &  (b)  &  (c)  &  (d) 
	\end{tabular}
	\end{center}
	\caption{ Feynman diagrams that contribute to the meson-mass at $O(p^4)$.
	Straight lines are the pseudoscalar meson propagator and wiggly lines are the photon.
	A filled dot represents an $\mathcal{L}^{(2)}$ vertex, while an open square represents an $\mathcal{L}^{(4)}$ insertion.
	(a) photon tadpole;
	(b) photon sunset;
	(c) meson tadpole;
	(d) $O(p^4)$ tree-level insertion.}
	\protect\label{fig:FeynDiag}
\end{figure}

\begin{figure}[b!]
	\begin{center}
	\begin{tabular}{c c c c c}
		\includegraphics[scale=0.1]{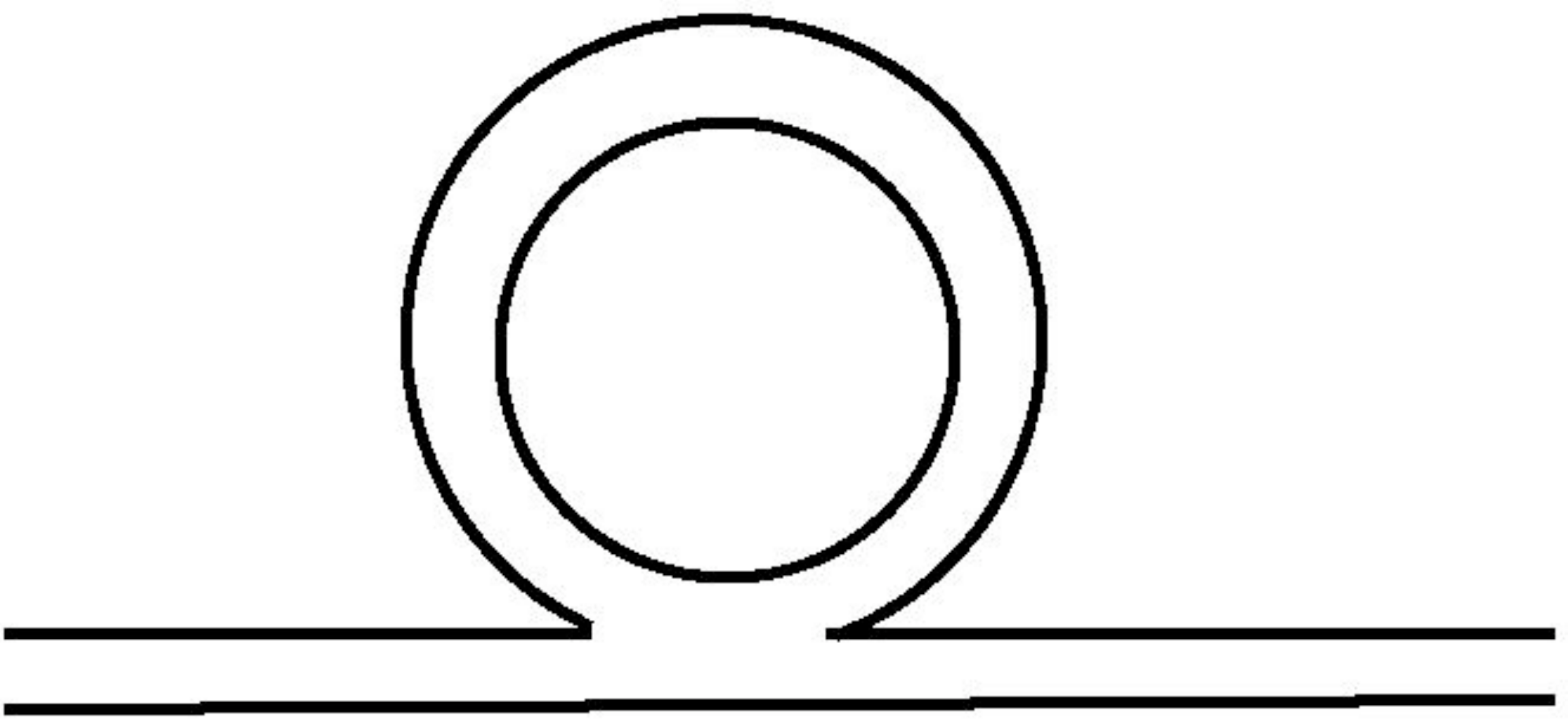}   % Connected
		&
		\includegraphics[scale=0.1]{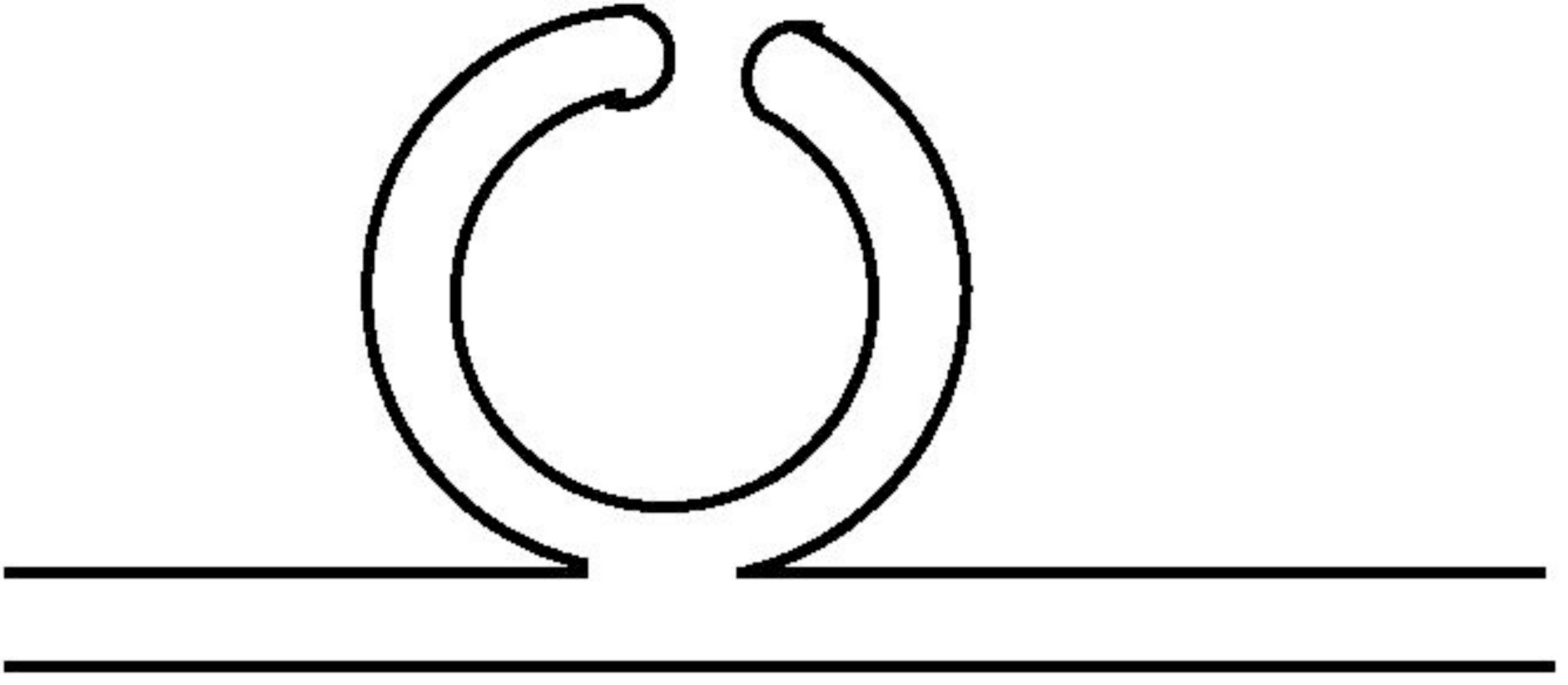}   % Hairpin
		&
		\includegraphics[scale=0.1]{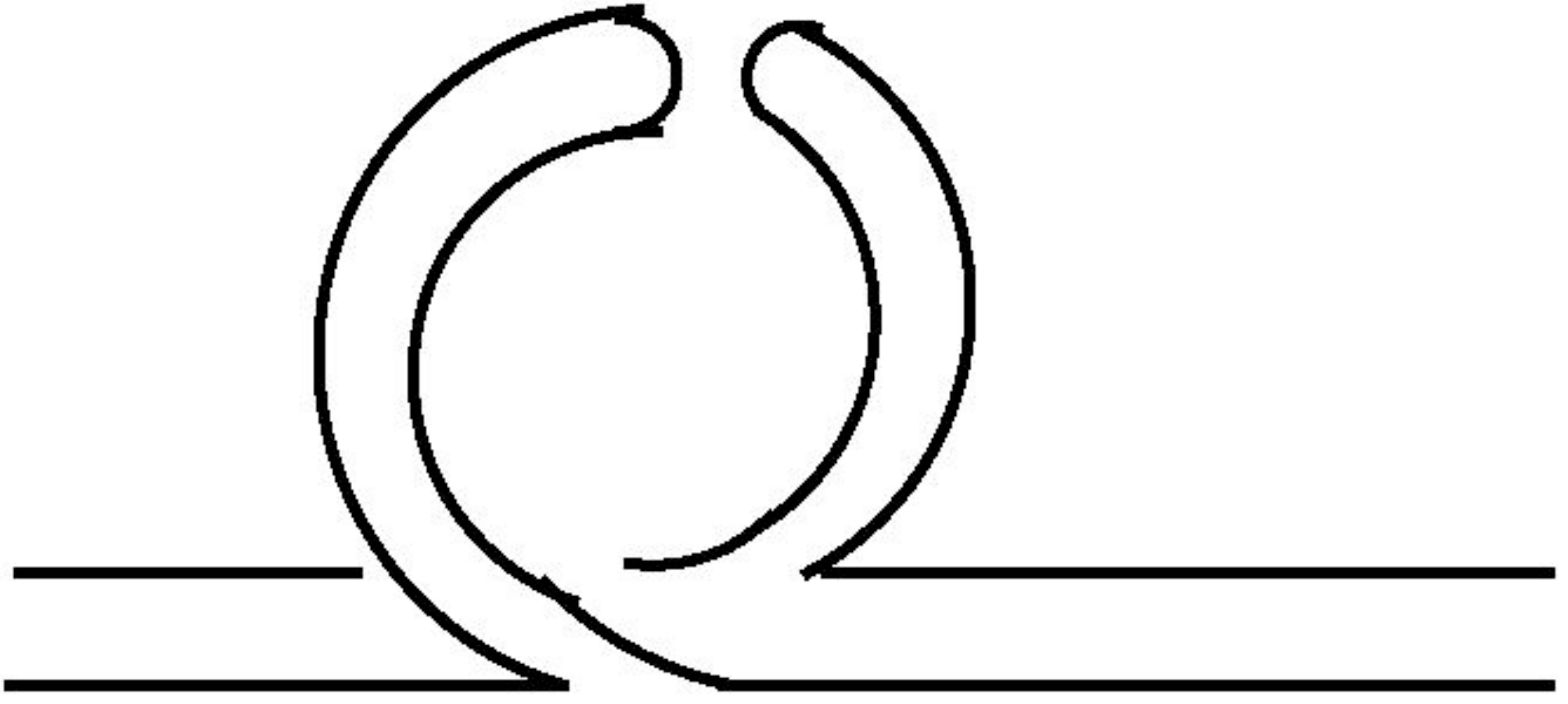}   % Fancy Hairpin
		&
		\includegraphics[scale=0.1]{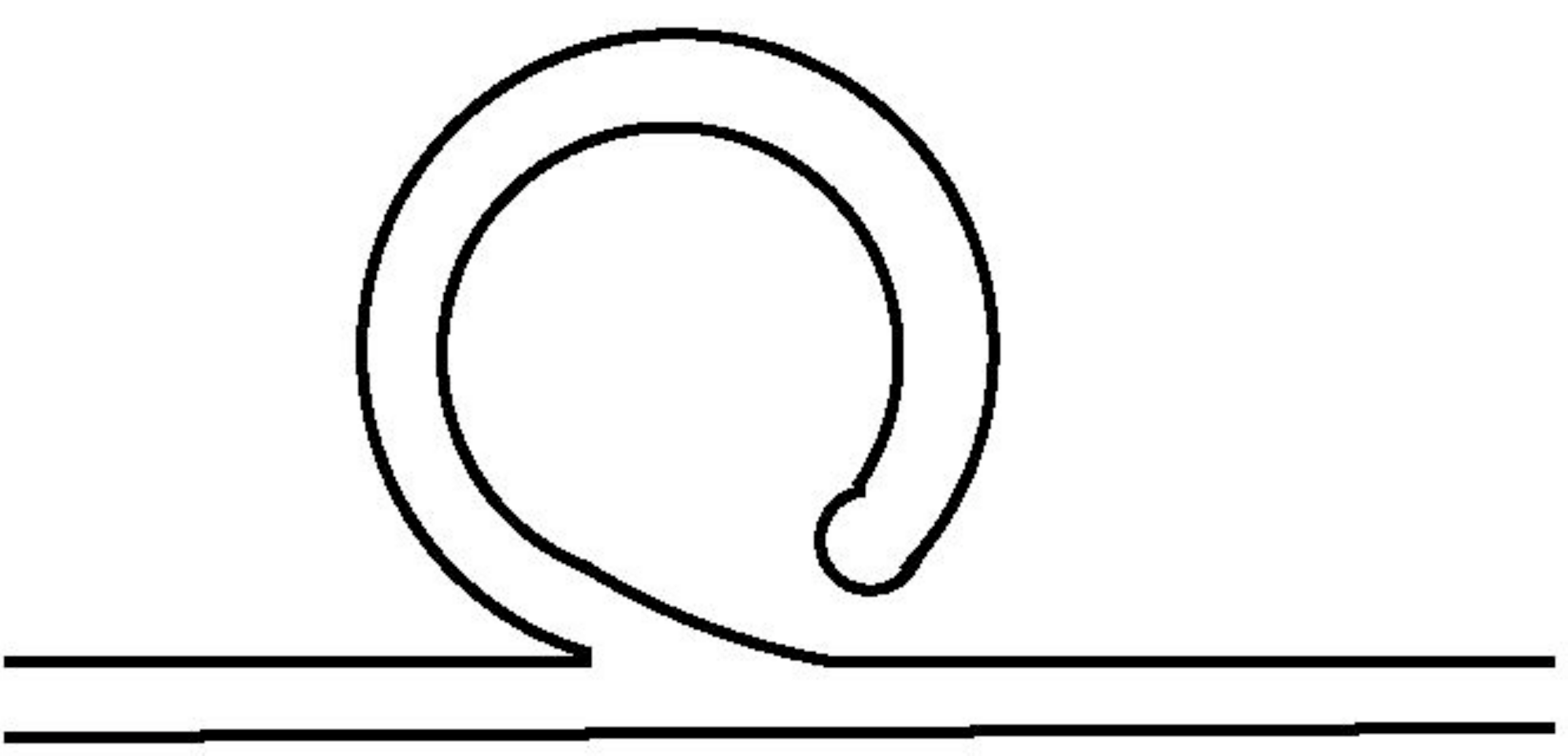}   % Coat Hanger
		&
		\includegraphics[scale=0.1]{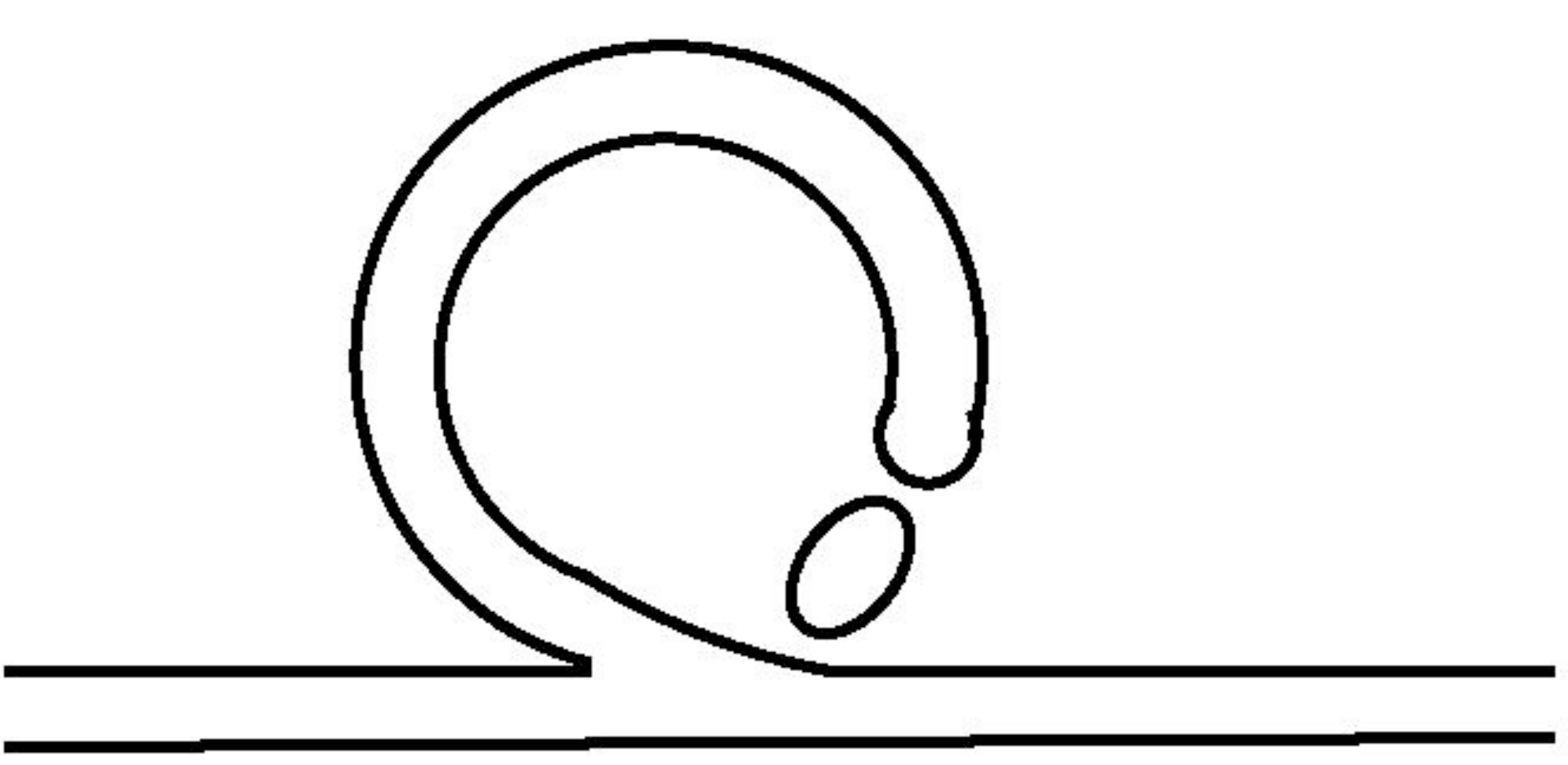}   % Coat Hanger + Bubbles
		 \\
		 (a)  &  (b)  &  (c)  &  (d)  &  (e)  
	\end{tabular}
	\end{center}
	\caption{ Quark flow diagrams corresponding to the tadpole diagram in  Fig.~\protect\ref{fig:FeynDiag}(c).
	(a) the diagram with both connected propagators and vertices: a ``connected'' diagram.
	(b) a disconnected propagator with connected vertices and a spectator quark: a ``hairpin''.
	(c) a disconnected propagator with connected vertices but no spectator quark: a ``fancy hairpin''.
	(d) a connected propagator with a disconnected vertex: a ``coat hanger''.
	(e) both disconnected propagators and vertex: the ``coat hanger + bubbles''.
	Note that any of the disconnected propagators can include bubble chains.  
	One sees a bubble explicitly in (e), because it is the leading-order, disconnected vertex + disconnected propagator diagram.}
	\protect\label{fig:quarkflow}
\end{figure}

Figure~\ref{fig:FeynDiag} shows the $O(p^4)$ contributions to the meson mass.
The photon tadpole contributes zero in dimensional regularization.
The photon sunset is straightforward to calculate, since the vertex is taste-conserving.
Calculating the contribution from the meson tadpole, Fig.~\ref{fig:FeynDiag}~(c), is facilitated by using quark-flow diagrams~\cite{Sharpe_quarkflow}.
Figure~\ref{fig:quarkflow} shows all contributing diagrams considering connected and disconnected propagators and vertices.

% FIGURE: Feynman Diagrams

% FIGURE: Quark Flow

The calculation closely follows that of Ref~\cite{Aubin:2003mg}, taking care to include \EM\
effects in the LO meson mass.
The end result is an \EM\ contribution from the connected diagram, and a non-\EM\ contribution from the disconnected diagrams that matches that of Ref.~\cite{Aubin:2003mg} if $m_{P_5^+}^2$ in their Eq.~(48) is replaced with $B_0 (m_x + m_y)$.
where $m_x$ and $m_y$ are the masses of valence quarks $x$ and $y$, respectively.%
\footnote{In Ref.~\cite{Aubin:2003mg}, $m_{P_5^+}^2 = B_0 (m_x + m_y)$.
Including electromagnetism, $m_{P_5^+}^2 = B_0 (m_x + m_y) + q_{xy}^2 \dem$.}

Specifically, for a meson comprised of quark $x$ and anti-quark $y$, the 
non-analytic \EM\ contribution is
\bea
	\delta M_{xy, 5}^2
	&=& - \frac{1}{16 \pi^2} \, e^2 q_{xy}^2 \, M_{xy,5}^2 \left[ 3 \ln (M_{xy,5}^2/\Lambda_\chi^2)   - 4  \right]   \nonumber \\
	 &&  \frac{-2 \dem}{16 \pi^2 f^2}  \left(\frac{1}{16}   \right) 
				  \sum_{\sigma,\xi}  \Big[   q_{x \sigma} q_{xy}  \,  
\ell(M^2_{x \sigma, \xi})
 				-   q_{y \sigma} q_{xy} \, \ell(M^2_{y \sigma, \xi})    \Big]  ,   \label{eq:result}
\eea
where $q_{xy} = q_x - q_y$, sea quarks are labeled by $\sigma$, the sixteen meson
tastes are labeled by $\xi$, $\Lambda_\chi$ is the chiral scale, and $\ell(M^2)$ is 
the renormalized loop integral 
\be
\int	\frac{d^4k}{\pi^2} \frac{1}{k^2 + M^2} \to \ell(M^2) \equiv M^2\ln(M^2/\Lambda^2_\chi)
\ee
The result in the first line in Eq.~(\ref{eq:result}) is from the photon sunset diagram, and that in the second line is from the meson tadpole. 

The contributions from Fig.~\ref{fig:quarkflow}~(d) lead to analytic contributions with unknown low-energy constants (LECs).
We concern ourselves here only with those which are new in the staggered-\EM\ calculation, namely the ones $O(a^2 e^2)$.
In principle, one could expect contributions from terms like $a^2 \sum_\sigma q_\sigma^2$ 
from the sea sector\footnote{The mixed-charge (cross-term) in the sea-sector is zero since $\langle Q \rangle=0$.}, 
and $a^2(q_x^2 + q_y^2)$, and $a^2 q_{xy}^2$ from the valence charges.
In practice, though, one only has terms $\propto a^2(q_x^2 + q_y^2)$ and $\propto a^2 q_{xy}^2$.

% THE USE OF QUENCHED-PHOTON SIMULATIONS
\section{Utility of Quenched-Photon Simulations}

As was first pointed out in
Ref.~\cite{Bijnens:2006mk}, \EM\ meson splittings
can be extracted at NLO in ChPT from quenched-photon simulations 
with no inherent systematic error due to quenching.
There are two places where sea-quark charges can appear: in the non-analytic (logarithm) terms and in the analytic (LEC) terms.
The log terms are completely calculable at NLO. Hence, we can put in by hand  the difference between the result of
a simulation with quenched photons (vanishing sea-quark charges) and the value with physical sea-quark charges.
For example, the contribution to Eq.~(\ref{eq:result}) coming from 
$q_\sigma\not=0$ can be added back in at the end
of the calculation.  This is not true for the 
analytic terms, which include, of course, unknown LECs.  
Nevertheless,
the only sea-charge dependent LEC terms are $O(m e^2)$ and take the form $\propto (m_x + m_y) \sum_\sigma q_\sigma^2$.
This means they drop out of the splittings $M_{K^\pm}^2 - M_{K^0}^2$ and $M_{\pi^\pm}^2 - M_{\pi'}^2$.

% CONCLUSIONS
\section{Conclusions}

We have discussed the origin of \EM\ effects in quark masses as they arise in a typical lattice calculation and shown that the lattice calculation of 
$M_{K^\pm}^2 - M_{K^0}^2$ and $M_{\pi^\pm}^2 - M_{\pi^0}^2$ can be used to improve the accuracy and precision of such calculations.
For improvement of the calculations in Ref.~\cite{MILC_quarkmass}, the MILC collaboration has begun staggered-quark, quenched-photon simulations~\cite{Torok}.
In these proceedings, we have described results of the rooted staggered \cpt\ calculation needed to extrapolate that lattice data.
We have also discussed why quenched-photon simulations are sufficient for calculating the needed meson-mass splittings.

% ACKNOWLEDGMENTS
\section{Acknowledgments}

This work was supported in part by the U.S. Department of Energy under Grant No.~DE-FG02-91ER40628 (C.B, E.D.F.)
and by the National Science Foundation under Grant No.~PHY-0555235 (E.D.F.).
This calculation was done in part using the Sage open-source mathematical software system~\cite{sage}.

% BIBLIOGRAPHY

\end{document}